\documentclass[12pt,preprint]{aastex}

\newcommand{\subsun}{\mbox{$_{\odot}$}}
\newcommand{\teff}{$T_{\rm{eff}}$}
\newcommand{\grav}{log($g$)}
\newcommand{\etal}{{\it et al.\/}}

\newcommand{\mystar}{HE1424$-$0241}

\begin{document}

\title{A New Type of Extremely Metal Poor Star\altaffilmark{1}}

\author{Judith G. Cohen\altaffilmark{2},  
Andrew McWilliam\altaffilmark{3},  
Norbert Christlieb\altaffilmark{4},
Stephen Shectman\altaffilmark{3}, Ian Thompson\altaffilmark{3}, 
Jorge Melendez\altaffilmark{5}, 
Lutz Wisotzki\altaffilmark{6} \& Dieter Reimers\altaffilmark{7} }

\altaffiltext{1}{Based in part on observations obtained at the
W.M. Keck Observatory, which is operated jointly by the California 
Institute of Technology, the University of California, and the
National Aeronautics and Space Administration.}

\altaffiltext{2}{Palomar Observatory, Mail Stop 105-24,
California Institute of Technology, Pasadena, Ca., 91125, 
jlc@astro.caltech.edu}

\altaffiltext{3}{Carnegie Observatories of Washington, 813 Santa
Barbara Street, Pasadena, Ca. 91101, andy, ian, shec@ociw.edu}

\altaffiltext{4}{Current address: Department of
   Astronomy and Space Physics, Uppsala University, Box 515,
   75120 Uppsala, Sweden, formerly at Hamburger Sternwarte, Universit\"at
Hamburg, Gojenbergsweg 112, D-21029 Hamburg, Germany, norbert@astro.uu.se}

\altaffiltext{5}{Palomar Observatory, Mail Stop 105-24,
California Institute of Technology, Pasadena, Ca., 91125,
Current address: Australian National University, Australia,
jorge@mso.anu.edu.au}

\altaffiltext{6}{Astrophysical Institute Potsdam, An der Sternwarte 16,
D-14482 Potsdam, Germany, lwisotzki@aip.de}

\altaffiltext{7}{Hamburger Sternwarte, Universit\"at
Hamburg, Gojenbergsweg 112, D-21029 Hamburg, Germany,
dreimers@hs.uni-hamburg.de}

\begin{abstract}

We present an abundance analysis for the extremely metal poor star \mystar\
based on high dispersion spectra from HIRES at Keck.
This star is a giant on the lower red giant branch with [Fe/H] $\sim -4.0$~dex.
Relative to Fe, \mystar\ has normal
Mg, but it shows a very large deficiency of Si,
with $\epsilon$(Si)/$\epsilon$(Fe) $\sim$ 1/10
and $\epsilon$(Si)/$\epsilon$(Mg) $\sim$ 1/25 that of all previously
known extremely metal poor giants or dwarfs.  It also has a moderately large deficiency of
Ca and a smaller deficit of Ti, combined with enhanced Mn and Co
and normal or low C. 
We suggest that in \mystar\ we see the effect of a 
very small number of contributing supernovae, and
that the SNII contributing  to the chemical inventory of \mystar\
were biased in progenitor mass or in explosion characteristics
so as to reproduce its abnormal extremely  low Si/Mg ratio. 
\mystar\ shows a deficiency of the explosive $\alpha$-burning elements
Si, Ca and Ti coupled with a ratio [Mg/Fe]
normal for EMP stars; Mg is produced
via  hydrostatic $\alpha$-burning.
The latest models of nucleosynthesis in SNII  fail to reproduce
the abundance ratios seen in \mystar\ for any combination of
the parameter space  of core-collapse explosions they explore. 

\end{abstract}


\section{Introduction}

Extremely metal poor stars provide important clues to the chemical
history of our Galaxy: the role and type of early SN, the
mode of star formation in the proto-Milky Way, and the formation
of the Galactic halo.  The  classes and properties of EMP stars are summarized by \cite{beers05}.
The peculiarities discussed there revolve around enhancements of the elements
C and N, which are often accompanied by enhancements of the
neutron capture elements beyond the Fe peak.
Mass transfer within a binary system which occurred while the former
primary was an AGB star is an explanation widely suggested for
the majority of these peculiarities, including excesses of both
CNO and heavy $s$-process neutron capture elements, see, e.g. \cite{cohen06}.

The number of extremely metal poor (EMP)
stars known below [Fe/H] $-3.5$ dex\footnote{The 
standard nomenclature is adopted; the abundance of
element $X$ is given by $\epsilon(X) = N(X)/N(H)$ on a scale where
$N(H) = 10^{12}$ H atoms.  Then
[X/H] = log$_{10}$[N(X)/N(H)] $-$ log$_{10}$[N(X)/N(H)]\subsun, and similarly
for [X/Fe].} is very small.  We have been
trying to increase it through
data mining of the Hamburg/ESO Survey (HES) \citep{wis00}. 
In this
paper we report our discovery of an extremely metal poor star which  
shows  peculiarities in its chemical abundance distribution
not seen in any other such star to date that is known to the authors.

\section{Stellar Parameters and Analysis}

\mystar\ (R.A.=14 26 40.3, Dec= $-02$ 54 28, J2000)
was observed in May 2004 with HIRES \citep{vogt94} at the 
Keck~I telescope.  Based on this high resolution spectrum, whose
total exposure time was 3600 sec, 
it was recognized at that time as an interesting EMP star with very low
Fe-metallicity.  It was observed again with HIRES in April 2006
after the detector upgrade with a total exposure time of 6000 sec.  This yielded
wider spectral coverage extending far into the UV and 
a better signal-to-noise ratio than the original data.

To determine stellar atmosphere parameters we use the procedures
described in \cite{cohen02} and adopted in all
subsequent work by our 0Z project published to date. 
Our \teff\ determinations are based on broad band colors
$V-I, V-J$ and $V-K$.
The IR photometry is taken from 2MASS \citep{2mass1,2mass2}.
We have
obtained new photometry 
at $V$ and at $I$ for \mystar\ ($V = 15.45\pm0.03$ mag and
$I = 14.54\pm0.03$~mag) from
ANDICAM images taken for this purpose over the past two years via a service
observing queue on
the 1.3m telescope at CTIO operated
by the SMARTS consortium\footnote{See http://www.astronomy.ohio-state.edu/ANDICAM and
http://www.astro.yale.edu/smarts.}.
We derive surface gravities through combining these \teff\ with
an appropriate
12 Gyr isochrone from the grid of \cite{yi01}.
We thus derive \teff = 5195~K and \grav = 2.50~dex.  The narrow Balmer
lines do not permit the star to be a dwarf below the main sequence turnoff.

The abundance analysis was carried out in a manner similar to those
described in \cite{cohen04}.  Full details will be
given in an upcoming paper which will present the most metal poor
stars we have found thus far.
If \teff\ for \mystar\ were to be increased by 100~K, the deduced [Fe/H] 
would increase by 0.15~dex, but the abundance ratios [X/Fe] would
be essentially unchanged.

\section{Abundances in \mystar}

The abundances we derive for \mystar\ are given in Table~\ref{table_abund}.
The number of lines used and the $\sigma$ of the derived 
log[$\epsilon$(X)] is given for each species for which absorption
lines could be detected; upper limits for some key
elements are included.  These results are compared to the
evaluation at [Fe/H] $-4.0$~dex of linear fits to the
abundance ratios determined by our 0Z project of
stars from the HES from the 0Z project (many still unpublished) 
with \teff $< 6000$~K and without substantial
carbon enhancement  ([C/Fe] $< +1.0$~dex).  
In the last column of the table we give same as determined
by \cite{cayrel04} for EMP giants.
The dispersion about their regression lines 
for giants with $-4.2 <$ [Fe/H] $< -3.1$~dex is 
small, only 0.11 dex for Mg, 0.20 dex
for Si, and 0.11 dex for Ca.  The extremely good agreement between
the abundance ratios for EMP stars found by these two independent large survey
projects, our 0Z project and the First Stars VLT project, 
and listed in the table is very gratifying, and provides support for our
statements about the extreme peculiarities of \mystar.

\begin{figure}
\epsscale{0.7}
\plotone{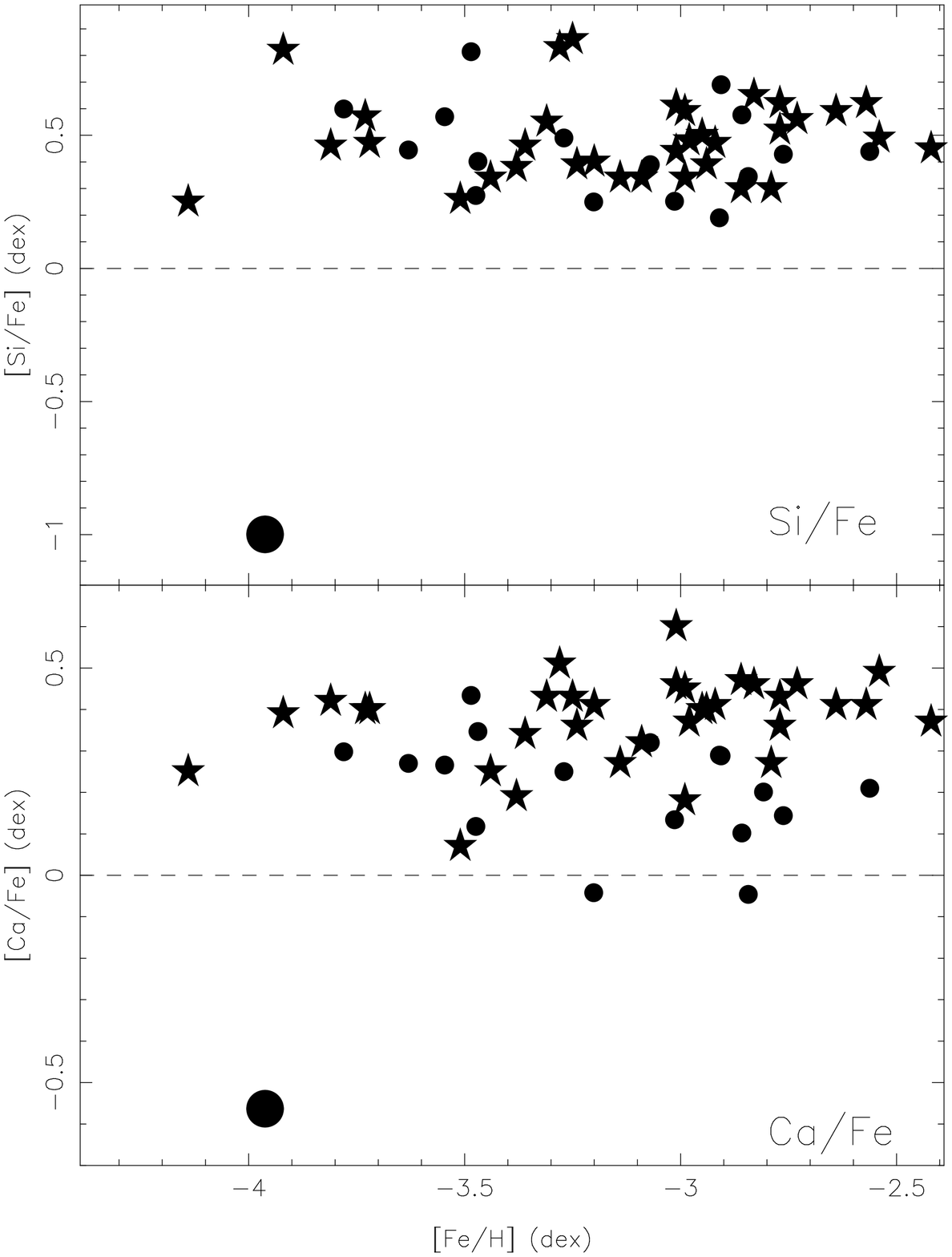}
\caption[]{[Si/Fe] (upper panel) and
[Ca/Fe] (lower panel) are shown as a function of [Fe/H] for EMP giants with
[Fe/H] $< -2.4$~dex.  Filled circles denote HES stars from our 0Z
project and star symbols are giants from the First Stars VLT project \citep{cayrel04}.
\mystar\ is shown  as the large filled circle, and is the only outlier,
being very low in both panels.  The dashed horizontal lines represent
the Solar abundance ratios.
\label{figure_sica_feh}}
\end{figure}

The anomalies seen in \mystar\ are many.  The most extreme and most
peculiar is the very large deficit of Si, with [Si/Fe] $\sim -1.0$~dex
and [Si/Mg] $\sim -1.4$~dex,
while all other known EMP stars have [Si/Fe] $\sim +0.3$~dex
and [Si/Mg] $\sim -0.3$~dex.
[Si/Fe] is low in \mystar\ by more than 6$\sigma$\footnote{$\sigma$ here is
the sum in quadrature of the uncertainty in [X/Fe] for \mystar\ 
and that of the uncertainty of the linear regression for the 
``normal'' EMP giants.}  compared to all
other known EMP giants, as is shown in Fig.~\ref{figure_sica_feh}.
\mystar\ also has a moderately large deficiency of
Ca (significant at the 5$\sigma$ level) and a smaller deficit of Ti.
It has enhanced Mn and strongly enhanced Co (significant at the 4$\sigma$ 
level), both odd atomic number elements.  Copper (another
odd atomic number Fe-peak element) may also be enhanced but
the single detected line is the rarely observed resonance line at 3274~\AA.
Carbon is not enhanced and the heavy neutron capture elements
Sr and Ba
have low abundances relative to Fe, suggesting that mass transfer
in a binary system involving an AGB star is not the cause of
the peculiar abundance ratios found in \mystar.
Each of these anomalies are seen in both the May 2004 and April 2006
HIRES spectra.  For example, the equivalent width of the only detected
Si~I line (at 3905~\AA)
is 17.7~m\AA\ from the 2004 spectrum
and 13.9~m\AA\ from the latter one.

No other EMP star shows the low Si/Fe and
Ca/Fe ratios seen in \mystar.
With one minor exception,
no other EMP dwarf or giant that is not C-enhanced is known to show
highly statistically significant abundance ratio deviations for any elements
between 
Mg and Ni.  (C-enhanced EMP stars sometimes show large enhancements
of the light elements, for example CS22949$-$037, found by
McWilliam et al 1995, analyzed again by Depagne et al al 2002.)
The exception is
the dwarf HE2344$-$2800 with [Fe/H] $\sim -2.7$~dex, 
found in the Keck Pilot Project
(Cohen et al 2002, Carretta et al 2002) to have an excess of Mn,
with $\epsilon$(Mn)/$\epsilon$(Fe) $\sim$ twice the prevailing
value among EMP stars.  This has been confirmed by a better HIRES
spectrum acquired in 2004; this dwarf also has a small excess of Ti relative to Fe.
A few C-normal EMP stars (CS22169$-$035 and CS22952$-$015,
for example, both of which are
included in Fig.~\ref{figure_sica_feh}), have slightly low $\alpha$-elements, but,
as the figure clearly illustrates, in 
no case do they approach the anomalies seen in \mystar.

\section{Comparison With Predicted SNII Yields}

At least several SN contribute to the chemical inventory of stars
with [Fe/H] $\gtrsim -3$~dex, and the
observed ratios of the chemical elements are determined by a
sum over an assumed initial mass function of predicted SNII yields.
SNIa and AGB stars also contribute at
still higher metallicity and later times.
But given the very low metallicity of \mystar, 
ejected material from only a very small number of core collapse SN 
are presumed to have contributed to the material in this star.  We must
therefore find a  model SNII whose predicted nucleosynthetic yields
match the abundance ratios seen in this star.
$^{28}$Si is formed 
largely in regions interior to where the bulk of
the $^{24}$Mg is produced, although of course nothing of
either of these species remains in the central region of the
SN, which is mostly $^{56}$Ni. Thus the details of the SN explosion
model are important in determining the Si/Mg ratio in the ejected
material.  We require a range in the ratio of $^{28}$Si/$^{24}$Mg
in the ejected material of at least a factor of 10 to reproduce
the behavior of both \mystar, with its strong deficit of
explosive $\alpha$-burning elements but normal Mg (from hydrostatic
$\alpha$-burning) and of all previously
known ``normal'' EMP stars.

The older  models of \cite{woosley}
are much more effective at reproducing the observed distribution 
of abundance ratios in \mystar.  Mg/Si production varies
by a factor exceeding 10 in these models, with Mg yields
highest at masses near 35~$M$\subsun, while Si yields
reach their maximum in SNII with lower progenitor masses 
near 20~$M$\subsun.  These yields can qualitatively reproduce the
behavior seen in \mystar.

However, none of the SNII models in the grids
recently calculated by \cite{chieffi} and by \cite{kobayashi}
comes close to
reproducing the abundance ratios among the $\alpha$-elements
seen in \mystar.  Both studies provide predictions
of explosive
yields for SNII progenitors with a wide range of initial masses
from 13 to 35 or 50~$M$\subsun\ with a wide range of metallicities.
They included an extensive network
of nuclear reactions.  For the  mass cut adopted in each of these
two studies, they each predict 
yields after the radioactive decays  
for $^{28}$Si/$^{24}$Mg whose range over the entire
set of model explosions does not
exceed a factor of two.  However, the two studies differ
in which mass range of SN progenitors produces larger
ratios of $^{28}$Si/$^{24}$Mg, \cite{chieffi} favoring
lower mass progenitors, while \cite{kobayashi} suggests
progenitor masses at the upper end of the range they consider.
In no case does the predicted production ratio  [Si/Fe] become
less than $-0.1$~dex.

The most recent predictions of nucleosynthesis yields
in SNII have undoubtedly been tuned to reproduce the
behavior of the previously known stars EMP stars with [Fe/H] 
reaching down to $\sim -4$ dex.
The abundance ratios we have derived for \mystar, however, demonstrate
that these models do not reproduce the full range of
the behavior of nucleosynthesis achieved in real SNII
and seen among the most peculiar of the the large sample of EMP stars
we have studied in the 0Z project.

Production of the odd atomic number elements Mn and Co  occurs 
through incomplete Si-burning for Mn and complete Si-burning for Co. 
\cite{kobayashi} point out that the odd-to-even ratio
among the Fe-peak elements depends on the mixing-fallback process, the
explosion energy, and the neutron excess $Y_e$.  While again
no published model can reproduce the large excess of Co and Mn relative
to Fe seen in \mystar, we must hope that some combination
of these parameters can be found that will accomplish this task.

\section{Behavior of The $\alpha$-Elements }

Differential analyses of large samples of stars within a small range 
of \teff\ in the thin disk of the Galaxy as compared
to stars in the thick disk such as those of \cite{edvard93},
\cite{bensby04}, and \cite{reddy06} have been able to achieve
very high precision.  These surveys have
demonstrated that the trends of [X/Fe] versus [Fe/H] are not identical
between the various stellar populations of the Galaxy.
But in such studies to date, all the $\alpha$-elements are believed to
have varied together and to show the same trends.

A few moderately metal-poor halo field stars have
been found that appear to be $\alpha$-poor; \cite{fulbright02}
suggests that  lower [$\alpha$/Fe] stars are found among those with high
space velocities with respect to the local standard of rest, while
\cite{stephens02} suggest such stars are associated with the outer halo.
The most extreme $\alpha$-poor stars, including that found by
\cite{carney97}, were reviewed by \cite{ivans}.  However, these
stars show depletions of Na, Al, Mg, Si and Ca with respect to Fe.
They are sufficiently metal-rich compared to \mystar\ that their
chemical inventory has a composite origin, with SNIa, SNII and AGB
stars all contributing, and can be qualitatively explained by
varying the SNIa/SNII ratio, an explanation which
cannot be applied to \mystar. 

All this, while interesting, is not the key issue for the abundance
distribution of the EMP giant \mystar. 
\cite{woosley} find that Si, Ca, and Ti are formed by explosive 
$\alpha$ burning in SNII, while O and Mg are produced by hydrostatic
$\alpha$ burning.  \mystar\ shows a clear large deficiency
of the former elements, but no apparent deficiency of the hydrostatic
$\alpha$ burning element Mg.

As abundance analyses have reached higher levels of accuracy
(or at least of internal accuracy) and as sample sizes have increased, 
there have been reports
of small differences, much smaller than those we find in \mystar,
between the behavior of the explosive and hydrostatic
$\alpha$-elements in certain specific environments.
The recent analyses of \cite{fulbright07} of a
sample of 27 red giants with Keck/HIRES spectra in Baade's Window 
in the Galactic bulge found
that the explosive $\alpha$-elements
Si, Ca and Ti have similar trends of 
[X/Fe] as a function of [Fe/H].   However they found
that the hydrostatic $\alpha$-elements
O and Mg show a different behavior in the bulge giants.  
This separation within the Galactic bulge sample is small;
$\sim$0.2~dex in total, much smaller than what we observe in \mystar.
\cite{fulbright07} detected similar effects, again on
a much smaller scale than in \mystar,
in a second environment,
among stars in the Milky Way dwarf spheroidal satellite galaxies.  They
used the compilation of data from the literature by \cite{venn}, which relies
heavily on the work of Shetrone, see, e.g. \cite{shetrone}.

These examples demonstrate that the production ratios of
the explosive to hydrostatic $\alpha$ elements are not fixed;
they must depend
on environment, the IMF, the star formation history, or other
relevant factors.  The subtle differences seen in the Galactic bulge 
and in dSph giants between the behavior of these two groups of
$\alpha$ elements, with the the explosive $\alpha$-elements
being more depleted than the hydrostatic ones, are seen in a much
more dramatic fashion in \mystar.
\mystar\ is a very extreme example of this phenomenon 
in a situation where only
a very few SN contributed to the chemical inventory of this star
and where, because of the very low metallicity of
\mystar, most other possible explanations for this become irrelevant.

\section{Summary}

All C-normal EMP giants studied to date in the two major surveys,
our 0Z project \citep{cohen04} and the First Stars VLT project \citep{cayrel04},
show smooth trends of abundance ratios [X/Fe] with Fe-metallicity with
modest dispersion around these trends and no strong outliers.
\mystar, with [Fe/H] $\sim -4.0$~dex, 
breaks this paradigm.  It is a many $\sigma$
outlier in several of the abundance ratios, 
with $\epsilon$(Si)/$\epsilon$(Fe) $\sim$ 1/10
and $\epsilon$(Si)/$\epsilon$(Mg) $\sim$ 1/25
that of all previously
known extremely metal poor giants or dwarfs, but normal
[Mg/Fe].  It also has a moderately large deficiency of
Ca and a smaller deficit of Ti, combined with enhanced Mn and
highly enhanced Co, both odd atomic number elements.  
With respect to Fe, C is normal or low in \mystar\
(the G band of CH was not
detected) and the heavy neutron capture elements are low.

From the point of view of SNII nucleosynthesis, \mystar\ is deficient
in the explosive $\alpha$-elements, but has a normal [Mg/Fe] ratio, where
Mg is produced in hydrostatic $\alpha$-burning.  Recent models
of production yields in SNII  fail completely to reproduce the behavior of
the $\alpha$-elements \mystar,
whose chemical inventory presumably resulted from a very small number
of previous SNII combined with any contributions from a hypothesized
Pop. III.  They also fail to reproduce the huge excess of Co with
respect to Fe.
These predicted yields are sensitive to the mass cut, the adopted electron
excess profile,  and to other explosion
characteristics assumed in the calculations for model SN.  They 
presumably were tuned to reproduce the behavior
of the previously known EMP stars, so their failure to come close
to reproducing the highly anomalous abundance distribution in \mystar\ is perhaps
understandable.

\mystar\ thus provides important clues as to the details 
of SNII explosions and their nuclear production  yields.
It is so metal poor that no explanation other than unusual core collapse
SN nucleosynthesis yields can be invoked to explain its unique
abundance ratios.
Modifications to standard
SNII models will need to be made to find explosion parameters which
can reproduce the properties we have derived for the peculiar
EMP giant \mystar.

\acknowledgements

We are grateful to the many people  
who have worked to make the Keck Telescope and HIRES  
a reality and to operate and maintain the Keck Observatory. 
The authors wish to extend special thanks to those of Hawaiian ancestry
on whose sacred mountain we are privileged to be guests. 
Without their generous hospitality, none of the observations presented
herein would have been possible.
This publication makes use of data from the Two Micron All-Sky Survey,
which is a joint project of the University of Massachusetts and the 
Infrared Processing and Analysis Center, funded by the 
National Aeronautics and Space Administration and the
National Science Foundation.
J.G.C. is grateful to NSF grant AST-0507219  for partial support.
      N.C. is a Research Fellow of the Royal Swedish Academy of
      Sciences supported by a grant from the Knut and Alice
      Wallenberg Foundation. He also acknowledges financial
      support from Deutsche Forschungsgemeinschaft through grants
      Ch~214/3 and Re~353/44.

{}

\clearpage

\begin{deluxetable}{l rrrrr | rr}
\tablewidth{0pt}
\tablecaption{Abundances for \mystar \label{table_abund}}
\tablehead{\colhead{Species} & \colhead{Log[$\epsilon$(X)]} &
\colhead{[X/H]}&  \colhead{[X/Fe]} & \colhead{$\sigma$} & \colhead{Number of} &
\colhead{[X/Fe](0Z)\tablenotemark{a}} &
\colhead{[X/Fe](VLT)\tablenotemark{b}} \\
\colhead{} & \colhead{(dex)} &
\colhead{(dex)} & \colhead{(dex)} & \colhead{(dex)} &
\colhead{Lines} & \colhead{(dex)} & \colhead{(dex)}  
}
\startdata     
C & $<5.26$ &   $<-3.33$ & $<+0.62$ & \nodata & CH  
 & \nodata & $\sim$+0.20 \\
N & $<5.10$ &  $<-2.83$ & $<+1.12$ & \nodata &  NH 
 & \nodata & \nodata \\
NaI\tablenotemark{c} &
    2.32 &     $-4.00$ & $-0.05$ & 0.07 & 2 & $-0.15$ & $-$0.20 \\
MgI & 4.03 &  $-3.51$ & +0.44 & 0.12 & 3 & +0.49 & +0.24 \\
AlI\tablenotemark{d} & 
    2.35 &     $-4.13$ & $-0.18$ & 0.15 &   2 & $-0.13$ &  $-0.12$ \\
SiI & 2.59 &  $-4.96$ & $-1.01$ & \nodata  & 1 & +0.45 &  +0.41 \\
CaI &  1.84 & $-4.52$ &  $-0.58$ & \nodata & 1 & +0.32 &  +0.27 \\    
CaII & 2.10 & $-4.26$ & $-0.31$ & \nodata & 1 & \nodata & \nodata \\
ScII & $-0.95$ & $-4.05$ & $-0.10$ & \nodata & 1  & +0.13 &  +0.04 \\
TiII & 0.85 &   $-4.14$ &   $-0.17$ & 0.17 &    8 & +0.29 &  +0.24 \\
VII & $<0.64$ & $<-3.36$ & $<+0.59$ &   \nodata & 1 & \nodata &
   \nodata   \\
CrI & 1.33 &  $-4.34$ & $-0.38$ & 0.09 & 5 & $-0.45$ & $-0.46$  \\
MnI\tablenotemark{e} & 1.59 &  $-3.80$ &   +0.15 &    0.02 & 2 &
     $-0.42$ & $-0.47$ \\
MnII & 1.69 &  $-3.70$ &   +0.25 &   0.13 & 2 &
  \nodata & \nodata \\
FeI & 3.49 &  $-3.96$ &   0.00 &    0.18 & 39 & 0.00 & 0.00 \\
FeII & 3.58 &  $-3.87$ &  +0.09 &    0.22 & 4 & 0.00 & 0.00 \\
CoI & 1.98 & $-2.94$ &   +1.01 &    0.21 & 4 & +0.50 & +0.40 \\
NiI & 2.52 &    $-3.73$ &    +0.22 &    0.01 & 2 & $-0.08$ & $-0.04$ \\
CuI & $-0.41$ &  $-4.62$ &   $-0.67$ &   \nodata & 1 & \nodata & \nodata   \\
SrII & $<-2.75$ &   $<-5.65$ &   $<-1.70$ &  \nodata & 2 
 & \nodata & \nodata \\
YII & $<-1.46$ & $<-3.70$ &  $<+0.25$ &    \nodata & 2 
  & \nodata & \nodata \\
BaII\tablenotemark{f} & $-2.74$ &   $-4.87$ &   $-0.92$ & \nodata & 1 
    & \nodata & \nodata  \\
EuII & $<-1.95$ &  $<-2.46$ &  $<+1.49$ & \nodata & 1 
     & \nodata & \nodata   \\
\enddata
\tablenotetext{a}{Regression lines for C-normal giants from our 0Z survey,
evaluated at $-4.0$~dex.}
\tablenotetext{b}{Regression lines from \cite{cayrel04}, Table~9, evaluated
at $-4.0$~dex.}
\tablenotetext{c}{Non-LTE correction of $-$0.2 dex has been applied for 
[Na/Fe] from the Na~D lines.}
\tablenotetext{d}{Non-LTE correction of +0.6 dex has been applied for 
[Al/Fe]  from the 3950~\AA\ doublet.}
\tablenotetext{e}{  The
adjustment of +0.4~dex for the 4030~\AA\ Mn~I triplet 
suggested by \cite{cayrel04} and by our own work has been applied here.}
\tablenotetext{f}{The Ba II line at 4554\AA\ is only
   marginally detected and therefore the Ba abundance may also
   be interpreted as an upper limit.}
\end{deluxetable}

\clearpage

\end{document}